# Blind Friendly Maps

## Tactile Maps for the Blind as a Part of the Public Map Portal (Mapy.cz)

Petr Červenka[1], Karel Břinda[2], Michaela Hanousková[1], Petr Hofman[3], Radek Seifert[4]

[1]Masaryk University, Support Centre for Students with Special Needs, Brno, Czech Republic
`{cervenka, hanouskova}@teiresias.muni.cz`
[2] Université Paris-Est Marne-la-Vallée, LIGM, Champs-sur-Marne, France
`karel.brinda@univ-mlv.fr`
[3] Seznam.cz, a.s., Prague, Czech Republic
`Petr.Hofman@firma.seznam.cz`
[4]Czech Technical University in Prague, Support Centre ELSA, Prague, Czech Republic
`radek.seifert@elsa.cvut.cz`

**Abstract.** Blind people can now use maps located at Mapy.cz, thanks to the long-standing joint efforts of the ELSA Center at the Czech Technical University in Prague, the Teiresias Center at Masaryk University, and the company Seznam.cz. Conventional map underlays are automatically adjusted so that they could be read through touch after being printed on microcapsule paper, which opens a whole new perspective in the use of tactile maps. Users may select an area of their choice in the Czech Republic (only within its boundaries, for the time being) and also the production of tactile maps, including the preparation of the map underlays, takes no more than several minutes.

## 1      Introduction

Acquiring proper spatial knowledge is fundamental for orientation and mobility of blind people and tactile maps are considered to be the most appropriate source of information about space [1]. However, one of the main obstacles to their use by the visually impaired is their problematic accessibility because they are not usually immediately available and updated. Therefore, relief embossing is still a result of demanding manual work and relatively costly and time-consuming procedures.

   The idea presented in this paper was born in 2007 when our team was working on ensuring accessibility of the web services provided by the Seznam.cz company. As a result of long-standing efforts, we developed a method of generating maps which are tactile and based on standard visual map underlays. The selected map underlay is consequently converted into a graphic document in conformity with the principles of tactile perception, and it is optimized for technologies using microcapsule paper.

Several authors have dealt with the issues of tactile map production over recent years. Watanabe [11] focused on possibilities of printing tactile maps from Open Street Maps source data – individual maps can be generated using the same microencapsulation technology and starting and finishing points may be added while no more explanatory notes are required. Authors have also explored ways of using 3D printing [9], which undoubtedly has its merits, yet is highly demanding in terms of technical equipment and financial resources. Interactive maps with audio-tactile content [4], [10] offer extra convenience. Nevertheless, it was our objective to describe a technological procedure that would be easily accessible to blind users and, at the same time, allow quick generation of updated tactile maps of any area. The result is the public map service www.hapticke.mapy.cz.

## 2 Possibilities and Methods of Automated Tactile Map Production

### 2.1 Points of Departure/Initial Assumptions

When preparing tactile maps, the designer needs to deal with specific issues related particularly to the principles of tactile perception [3], [5]. The designer must:

1. Determine the optimal ratio between the rendered content (maximum of rendered details) and the size of the resulting image (readable through touch), i.e. the **map format** and its **scale**.

2. Reduce significantly the **map content** and select unambiguously individual elements to be shown on the map (point, line and area symbols).

3. Choose a suitable way of **map lettering**.

4. Choose the relief embossing **technology**.

All these aspects significantly influence each other (different technologies enable rendering different number of details) and a suitable compromise needs to be found. In addition, it is necessary to bear in mind other conditions of legibility of the map with regard to the limits of tactile perception. These conditions often include a unified size of Braille characters (offering no possibility of multiple sizes) and the need to maintain sufficient distance between the individual elements of tactile image [1], [5]. The described approach was consistently applied when creating our map key— the Mapnik, a freely accessible map generating interface, was used [12].

### 2.2 Map Scale Selection

The basic map scale was derived from the minimum width of the road which allows the Braille lettering to be rendered in the correct size within its axis. The basic format of the map sheet is the A4 swell paper size.

### 2.3 Rendering and Map Key Generation

The difference between generating standard and tactile maps using the Mapnik program is the unequal use of the configuration file, the so-called map key. The map key defines objects to be chosen from the SQL vector database and determines the order and manner of their rendering. The technical nature also imposes certain limitations applying to the final shape of the map. These can be divided into two groups:

a) Limitations imposed by the employed map generating tool, that is, the aforementioned Mapnik interface, are due to its possibilities of rendering and the set of rules that can be used. For instance, you cannot change the location of an object on the basis of the presence of other objects in its vicinity. This might lead to situations when e.g. a brook disappears under the path running alongside.

b) The second category of limitations are those imposed by the chosen object database. This often results from the fact that the database is primarily created to serve for standard maps rendering, not as a generally accurate representation of all rendered objects. That explains, for instance, why bridges are stored in the database only as a line of both bridge decks which are rendered correctly in the conventional maps, but cannot be used when rendering tactile maps. The distance between them is far too small and their depiction would thus render the neighboring objects worthless.

In addition, it was necessary to select elements to be rendered from a large database, which contains several hundreds elements, and determine an appropriate level of map content generalization. The main method applied was the generalization method of elements classification and simplification, which allows to view the map content in a readable and clearly arranged format suitable for tactile perception.

The following chart summarizes the development of the resultant map from the geoinformation database:

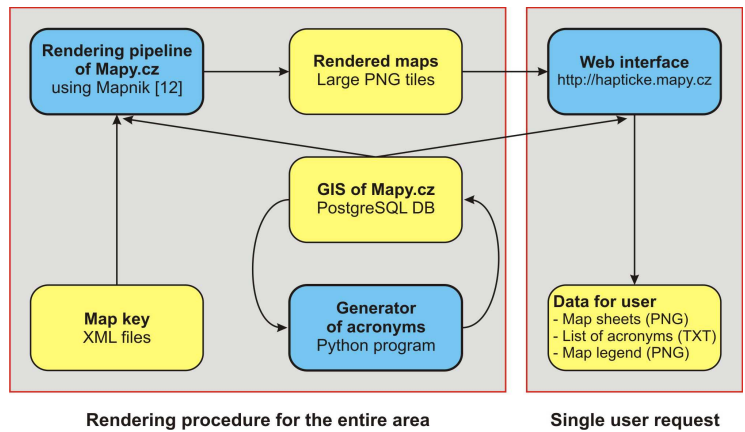

### 2.4 Testing

A user-based testing procedure has been carried out in order to choose fillings that may be unambiguously identified through touch both in isolation and in the context of a real map sheet. Seven blind students, all experienced users of tactile graphics on swell paper, were involved in the testing phase and undertook the assigned tasks. The four types of filling emerged from this testing phase. In the second phase of testing, the testers were given a complete map sheet. This time, the aim was to check whether the identification of line and area symbols was unambiguous and also to examine the whole conception, tactile readability, and clear arrangement of the map sheet.

### 2.5 Map Lettering

To ensure maps are lettered clearly in Braille system, an algorithm was developed. It generates abbreviations of street names and residences so that they are unique in the given area and at the same time intuitive. Explanatory notes of the abbreviations located in the selected area are then exported to an individual text file. It is necessary to ensure sufficient distance of the description from the adjacent drawing; in case an overlap is inevitable, it is imperative to give the description an absolute priority.

### 2.6 Print Technology

We opted for the swell paper technology as it is the most widely available technology of relief embossing in the Czech Republic, which means that users can obtain the generated maps more easily in comparison to other technologies.

## 3 Results

### 3.1 Map Sheet Rendering

Unlike conventional maps available on web-based map servers, map underlays are rendered on individual, predefined map sheets using a unique map sheet designation. The reason is that they are offered to the users as hard (tactile) copy which requires a fixed format. Individual sheets in A4 format render an area of 300 × 425 m at an approximate scale of 1:1500 (see Fig. 1).
The content of individual map sheets is based on the same source database as standard maps and has been processed with the same tool, that is, the aforementioned Mapnik [12]. The indisputable advantage of this practice is that all updates in the database are also immediately reflected in the newly generated tactile maps.

### 3.2 Map Key

**Area symbols.**
All areas were grouped as follows (see also Fig. 2): buildings, water bodies, green areas, industrial areas.

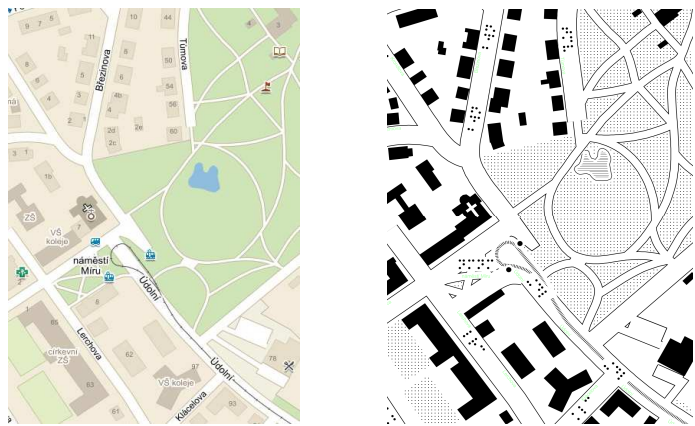

**Fig. 1.** Example of how an area is displayed on a standard and a tactile map

Buildings are marked as raised surfaces in their authentic ground plan. Water surfaces are marked using horizontal raster with an indented boundary line. The symbol for green areas applies to all grass and wooded areas; industrial areas are marked using a structured raster with an indented outline. All other areas are rendered without a raster to allow rendering other elements of the map.

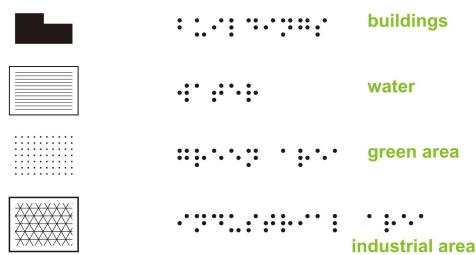

**Fig. 2.** Example of area symbols

**Line symbols.**
Streets and routes are the basic line elements of a tactile map. They are rendered in two different widths; the basic width option is wide enough to allow Braille map lettering (13 mm) while the narrow option (5 mm) is reserved mainly for paths in green areas and streets in high building density areas.
Other line symbols that complete the space and are thus rendered as well are tram lines, railways, and stairs. Other complementary line symbols are stand-alone walls, brooks, and cableways (see Fig. 3).

**Point symbols.**
Although the database of point symbols offers an inexhaustible number of possible objects, it was necessary, for the sake of comprehensibility and clarity of the render-ing, to reduce significantly the point symbols to merely three: tram stops, car parks, and churches located in blocks of houses.

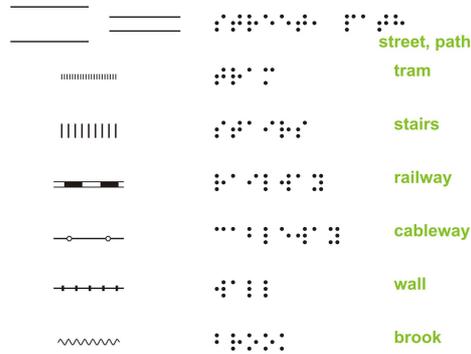

**Fig. 3.** Example of line symbols

### 3.3 Map Lettering

The map lettering is implemented in two ways in order to allow cooperation between blind and sighted users who cannot read Braille. The streets are lettered in the first place using three letters of their name – they are written in Braille in the street axis. Abbreviations are generated in a manner that prevents repetitions in the selected area. These descriptions are completed with full names of the streets in light green color. Letterings of the names of squares are produced similarly but four letters are used to aid recognition (see Fig. 4). All abbreviations, together with an explanatory note on full names, are generated into an auxiliary file which is part of the map.

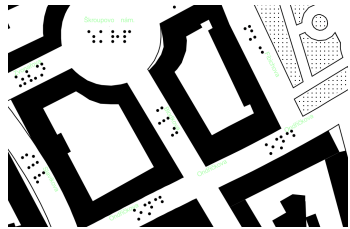

**Fig. 4.** Example of map lettering (streets are marked using the first three letters of their name in the street axis whereas squares are marked using the first four letters in Braille in the center)

### 3.4 Web Interface

**Map selection.**
Selecting the map of a given area (or a set of map sheets) does not differ from conventional map browsing, that is, searching for a location by its address, entering the name of the street (displaying the middle of the street), or typing the exact coordinates of the wanted point on the corresponding place of the map portal [14].

After the user has entered the desired location, the site generates a basic map sheet containing the requested location and the user can select the range of print (number of map sheets – see Fig. 5). Blind users may specify the print range using the spreadsheet mode (marking adjacent sheets). The site maintains the continuity between individual sheets to make generating larger maps possible. The map sheets can be complemented by map designation while minimizing interference with the map drawing.

In the next step, the user can download the archive with the separate sheets (graphic files in PNG format), the list of abbreviations (a TXT text file), and—in case the user has selected this option—even the map legend in graphic form (a PNG file).

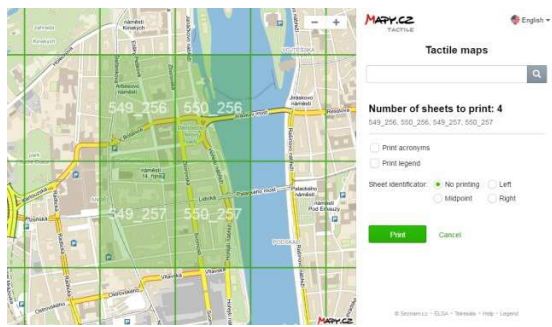

**Fig. 5.** Example of selecting map sheets for print

**Map printing.**
As mentioned above, the whole method uses swell paper technology. It is a relatively quick method which uses standard printing in the first stage. The content of the map is printed on a special sheet of paper and, after being processed by a special device equipped with an infrared lamp, the black print elements become embossed in the final stage. Currently we have at our disposal the following devices: the Zy Fuse Heater by the company Zychem and the P.I.A.F. by the company HARPO.

## 4    Discussion

An undeniable advantage of the discussed service is the possibility of generating tactile maps within a short period of time. Its availability contributes to the effective promotion of spatial orientation and development of tactile perception of the blind, while also raising public awareness. Finally, it facilitates the work of educational institutions.

However, two conditions must be met in order to enable a wider use of tactile maps: first, users need to consider them useful; second, they have to be able to interpret correctly the relatively complex map drawing. This requires either having experience with tactile map reading, or resolution and patience to acquire the needed skills.

Technically, it is necessary to consider the possibility of expanding the input geographic data by adding the output text attachments or implementing the content directly into the generated maps.

# 5   Conclusion

The service currently offers continuous view of the entire country, in a scale suitable for rendering urban environments with an emphasis on street network. In the future, we aim to achieve rendering of smaller scales and to increase the range of the rendered territory, as suggested below:

1. Smaller scale map rendering. Recently, we have prepared the map keys for tactile maps in scales of 1:100 000 and 1:50 000 (two basic levels of urban areas maps and a net of roads and railways).

2. A world map. We hope to prepare a basic map of the world using the Open Street Map data integration. This way, we would be able to offer a tactile map of any place in the world (depending only on the OSM data availability).